\documentclass[conference]{IEEEtran}
\IEEEoverridecommandlockouts
\usepackage{cite}
\usepackage{amsmath,amssymb,amsfonts}
\usepackage{algorithmic}
\usepackage{graphicx}
\usepackage{textcomp}
\usepackage{xcolor}
\usepackage{hyperref}
\usepackage{booktabs}
\usepackage{tikz}
\usepackage{flushend}
\usepackage{multirow}
\usepackage[colorinlistoftodos,prependcaption,textsize=tiny]{todonotes}
\usepackage[capitalise]{cleveref}
\usepackage{csquotes}
\def\BibTeX{{\rm B\kern-.05em{\sc i\kern-.025em b}\kern-.08em
		T\kern-.1667em\lower.7ex\hbox{E}\kern-.125emX}}

\newcommand{\drawPercentageBar}[1]{%
	\begin{tikzpicture}
		\fill[color=black] (0.0 , 0.0) rectangle (#1*5ex , 1.7ex );
		\fill[color=lightgray] (#1*5ex , 0.0) rectangle (5.0ex, 1.7ex);
	\end{tikzpicture}%
}

\newcommand{\summary}[2]{%
	\vspace{-0.2cm}%
	\begin{center}%
		\colorbox{gray!20}{%
			\parbox{\linewidth}{%
				\textbf{\textsf{Summary (\textit{#1})}:}~%
				#2%
			}%
		}%
	\end{center}%
}

\newcommand{\suggestion}[2]{
	\vspace{2mm}
	\noindent
	\fbox{%
			\parbox{.97\linewidth}{%
				\textbf{#1.}
				#2
			}%
		}%
		\vspace{2mm}
	}%

\begin{document}
		
		\title{A Survey on What Developers Think About Testing}

		\author{\IEEEauthorblockN{Philipp Straubinger}
			\IEEEauthorblockA{\textit{University of Passau} \\
				Passau, Germany}
			\and
			\IEEEauthorblockN{Gordon Fraser}
			\IEEEauthorblockA{\textit{University of Passau} \\
				Passau, Germany}
		}
		
		\maketitle
		
		\begin{abstract}
			Software is infamous for its poor quality and frequent occurrence of bugs.
			While
			there is no doubt that thorough testing is an appropriate answer to
			ensure sufficient quality, the poor state of software generally suggests
			that developers may not always engage as thoroughly with testing as
			they should. This observation aligns with the prevailing belief that
			developers simply do not like writing tests.
			In order to determine the truth of this belief, we conducted a comprehensive survey
			with 21 questions aimed at (1) assessing developers' current engagement with testing
			and (2) identifying factors influencing their inclination toward testing; that is,
			whether they would actually like to test more but are
			inhibited by their work environment, or whether they would really
			prefer to test even less if given the choice.
			Drawing on 284 responses from professional software developers, we uncover reasons
			that positively and negatively impact developers' motivation to test.
			Notably, reasons for motivation to write more tests encompass not only a general pursuit of software
			quality but also personal satisfaction.  However,
			developers nevertheless perceive testing as mundane and
			tend to prioritize other tasks.
			One approach emerging from the responses to mitigate these
			negative factors is by providing better recognition for developers' testing efforts.
		\end{abstract}

		\begin{IEEEkeywords}
			Motivation, Survey, Software Testing, Software Engineering, Empirical Study
		\end{IEEEkeywords}
		
		\section{Introduction}
		\label{sec:intro}

		Testing plays a pivotal role in the software development process, serving
		as the primary technique to ensure or improve software quality. If testing
		is not given due consideration from the	outset, the resulting software applications
		can suffer from fundamental faults and failures, leading to a bad user experience,
		crashes, incorrect computations, flawed data, and even project failure.
		This concern is especially pertinent given the
		sheer scale of today's large software projects with hundreds of
		thousands of lines of code in companies and open-source. Regrettably, the existence
		of software quality problems~\cite{pooreport} suggests that
		testing is inadequately applied in practice.

		There has been much research and speculation about the reasons for
		this. Notably, it has been observed that the role of a ``software tester'' is not
		always viewed as desirable, as testers often receive limited recognition within companies,
		and testing tasks are commonly regarded as challenging, with outcomes that
		are much less tangible than those achieved when writing code~\cite{DBLP:journals/software/WeyukerOBP00,DBLP:journals/corr/WaychalC16,DBLP:journals/infsof/DeakSS16}.
		Testing should thus not only be the responsibility of
		dedicated testers, but it would be preferable for it to also
		be an integral part of the work of software developers, who
		should not only implement but also test
		features~\cite{spillner2021software,DBLP:conf/esem/SantosMCSCS17}.
		Unfortunately, evidence suggests that developers tend to
		engage much less with testing than one might hope
		for~\cite{DBLP:journals/tse/BellerGPPAZ19}.

		The lack of developer testing raises the question: Are developers
		hindered by their work environment, such as time constraints or
		inadequate testability, do they deliberately choose not to test
		because they do not enjoy doing it, or do they maybe not have the skills and
		training to perform better testing? To shed light on these questions we
		conducted a survey targeting developers in practice, aiming to understand
		their testing habits and aspirations.
		%
		%
		%
		Using data collected from 284 respondents to a comprehensive survey comprising 21
		questions, we first aim to understand how developers perceive their engagement with testing, the nature of their testing efforts, and whether they
		feel these efforts receive appropriate recognition. Consequently, our first
		research question is as follows:
		
		\vspace{0.2em}
		\noindent\textit{\textbf{RQ 1}: How do developers engage with testing in practice?}
		\vspace{0.2em}
		
		\noindent Our findings indicate that developers often feel they have
		insufficient time for testing, resulting in inadequate testing efforts.
		Furthermore, testability emerges as a practical challenge, leading to
		intricate tests that are cumbersome to create. Despite these
		factors, developers express satisfaction with the adequacy of their testing
		efforts in over half of the projects.

		To gain a deeper understanding of the role of developers' motivation in their
		testing practices, allowing us to differentiate whether they would actually like
		to test more but are inhibited by their work environment, or whether
		they would prefer to test even less, our second research question is:
		
		\vspace{0.2em}
		\noindent\textit{\textbf{RQ 2}: How would developers like to engage with testing?}
		\vspace{0.2em}
		
		\noindent The responses reveal that developers are split regarding their desire to test more or less:
		On one hand, the majority express either a reluctance to increase testing or a preference to invest less time in it, often citing testing as a mundane task or deeming it less important than other responsibilities.
		On the other hand, a substantial number of developers express a desire to engage in more testing if it were technically easier or, notably, if their efforts received greater recognition from management and peers.
		Some developers are satisfied with their current testing, either having reached a quality goal or believing that further effort would provide no benefits.

		In summary, the contributions of this paper are as follows:
		\begin{itemize}
			\item We present a comprehensive survey 
			aimed at assessing the prevailing and desired testing behaviors of
			developers.
			\item We quantitatively and qualitatively evaluate the data,
			providing insights into developers' thoughts and aspirations.
			\item With the evaluated data, we can strengthen existing research
			and provide new insights into current problems and misbehaviors in
			software testing.
		\end{itemize}
		
		Overall, our results suggest that, while technical and organizational
		issues pose inhibitive factors, motivation significantly influences
		developers' testing practices. On the one hand, this reinforces
		ongoing research on automated testing and test generation to
		relieve developers of some of the challenges but on the other hand
		this suggests that future research needs to look into increasing the
		motivation of developers for testing.

		\section{Background} \label{relatedwork}
		
		\subsection{Developer Testing}
		
		Testing is usually done either by external testers (e.g.,
		quality assurance departments or companies), or by
		developers while coding. Both are established practices in
		companies~\cite{myers2011art}, but in this paper, we focus on
		developers.
		
		Depending on the current state of development of an application,
		developers may apply different approaches for testing, and these may
		lead to different challenges, perceptions, and motivations. During
		the early stages of the product or of individual features, unit tests target
		the smallest components of the software (e.g., methods or
		classes). Developers may also write automated integration tests,
		checking the interactions of integrated components and their
		interfaces. Finally, automated end-to-end tests are a common means to
		check the functionality and satisfaction of requirements at the system
		level~\cite{SpillnerLinz12}.
		
		Independently of the type of tests, a substantial gap between writing
		production code and test code has been observed~\cite{10.1145/2786805.2786843,DBLP:journals/tse/BellerGPPAZ19}, with developers spending
		about 75\,\% of their time for writing code and only about 25\,\% on
		testing. Interestingly, developers tend to overestimate the time used
		for writing tests, often claiming a 50 to 50 ratio, while in truth some developers do not test at
		all~\cite{10.1145/2786805.2786843,DBLP:journals/tse/BellerGPPAZ19}, sometimes resulting in entire projects without automatically executed tests. In this paper, we aim to
		understand whether this is caused by technical or organizational
		challenges, or rather just a lack of motivation.

		\subsection{Motivation and Engagement in Software Engineering}

		We aim to investigate factors of developer motivation 
		while testing, but the term ``motivation'' is overloaded, and we thus
		need to consider several different dimensions in our analysis.
		In everyday language, the terms motivation and satisfaction are used
		synonymously, but according to the new theory of work motivation and
		job satisfaction of software engineers~\cite{francca2014theory,
			DBLP:conf/esem/FrancaSS14} they are not the same: Motivation needs to
		be awakened before the work starts, while satisfaction is caused by results.
		Motivation and satisfaction are connected because
		being satisfied by the previous task can motivate the developer for
		the next task. Motivation can be intrinsic, referring to the inner
		willingness to do an activity for personal satisfaction, or extrinsic, referring to a separable outcome that comes with or after completing a
		task, such as recognition for a person's work \cite{ryan2000self}.
		The term engagement in the context of software engineering is defined
		as commitment, hard-working, and interest in the person's current work, which
		might go beyond the simple motivation to satisfy a task~\cite{8370133} and instead invest extra effort exceeding what is required~\cite{suff2008going, devi2009employee}.
		
		In our study, we focus on motivation as the initial factor for developers, as both satisfaction and engagement can only be achieved once they are motivated to test.
		While factors influencing motivation have been observed for software engineering in general~\cite{francca2014theory,
			DBLP:conf/esem/FrancaSS14}, it is not known yet whether these also apply to software testing.

		\section{Survey Design} \label{researchdesign}
		
		To answer RQ1 on the current and RQ2 on the desired engagement with testing (see \cref{sec:intro}), we designed a survey based on the
		guidelines by Lin{\aa}ker et
		al. \cite{8ac54dbeb7ac42449c430f0d157efa26}.
		
		\subsection{Questions} \label{sec:questions}
		
		The survey consists of a total of 21 mandatory questions and one optional one, divided into four categories. The questions are either single or multiple-choice questions, with or without an `others' free-text option, Likert scale choices, and stand-alone free-text questions.
		Designing a survey involves a trade-off between asking many questions and the resulting difficulty in acquiring survey responses and their costs. We chose and refined questions through an iterative process, where we based them on our research questions (RQs) and refined them through multiple steps within our research group and a pilot study involving different researchers. We tried to keep textual explanations brief and clear to ensure valid responses. This resulted in the revised questions shown in \cref{tab:allquestions} and their answer options included in the artifacts (\cref{conclusion}).
		
		The survey starts with demographic questions. Since we use Prolific (cf. \cref{sec:prolific}),
		an established provider of survey respondents, we do not require questions about information already provided by Prolific, in particular age, country of residence, employment status, sex, and student status. Beyond this, we ask the questions with the ID \enquote{UD} listed in \cref{tab:allquestions} regarding the participant's degree (UD1), experience (UD2), number of employees (UD3), and role in the company (UD4).
		
		The second category of the questionnaire consists of questions about the software projects the respondents work in (\cref{tab:allquestions}, IDs with PD). In particular, we query context information such as the project size (PD2, PD3), the working domain (PD1), and quality metrics used in the participant's projects (PD4--6). 
		Together, the demographic and project questions provide context for the testing-related questions.
		
		The third category asks about the current state and efforts for testing in the respondents' projects to answer RQ1. 
		In particular, we ask developers about their daily test behavior (CS1--3), recognition (CS4), and struggles while testing (CS5--7 in \cref{tab:allquestions}). These questions are of special importance for RQ2 likewise, since they serve as a baseline to see the deviation between desires and current testing practices in companies.

		In order to learn how the participants would like to test (RQ2) and what they want to change (DS1--4), questions with ID \enquote{DS} in \cref{tab:allquestions} enable direct comparison to the previous set of questions (IDs with UD). An additional optional question at the end asks whether participants have any other information to share about testing in their company that has not been sufficiently covered by the previous questions (AE1).
		
		\begin{table}[t]
			\caption{Questions of the pre-study and the main study \\ \tiny{with Single Choice as SC and Multiple Choice as MC}}
			\label{tab:allquestions}
			\resizebox{\linewidth}{!}{%
				\begin{tabular}{lll}
					\toprule
					ID & Question & Type     \\ \midrule
					\addlinespace[0.5em]
					\multicolumn{2}{l}{Questions in the pre-study} \\ \cmidrule(r){1-2}
					PS1 & Do you professionally develop/test software?       	& SC   \\ 
					PS2 & Do you currently work on a software project?       	& SC   \\ 
					PS3 & Do you write code or tests?         				& SC   \\ 
					\addlinespace[0.5em]
					\multicolumn{2}{l}{Questions in the category user demographics} \\ \cmidrule(r){1-2}
					UD1 &What is your highest type of graduation?                     & SC   \\ 
					UD2 & How many years of professional experience do you have?       & SC   \\ 
					UD3 & How many full-time employees does your company have?         & SC   \\ 
					UD4 & What is your current role in the company?                    & SC + free-text   \\ 
					\addlinespace[0.5em]
					\multicolumn{2}{l}{Questions in the category project demographics} \\ \cmidrule(r){1-2}
					PD1 & In which domain is your current project?                     		& SC + free-text   	\\ 
					PD2 & How many employees are working in your current project?      		& SC   	\\ 
					PD3 & What is the size of your current team?         						& SC   	\\ 
					PD4 & What are the metrics used for measuring quality in your project?    & MC + free-text   \\ 
					PD5 & What is the overall code coverage in your project? 					& SC 	\\
					PD6 & Is there a requirement for quality in the project? 					& SC + free-text 	\\
					\addlinespace[0.5em]
					\multicolumn{2}{l}{Questions in the category current situation} \\ \cmidrule(r){1-2}
					CS1 & How much time of your daily work do you use for testing?               	& SC   	\\ 
					CS2 & What kind of tests do you write/perform?	     						& MC + free-text  	\\ 
					CS3 & Do you prefer to test or write code?         							& Likert 5 point   	\\ 
					CS4 & What kind of recognition do you get for writing tests?    				& MC + free-text   \\ 
					CS5 & In your opinion, has your current project been tested well enough? 		& SC + free-text 	\\
					CS6 & How much effort does it take to write tests for the project? 			& Likert 5 points 	\\
					CS7 & How complex are your tests?												& Likert 5 points \\
					\addlinespace[0.5em]
					\multicolumn{2}{l}{Questions in the category desired situation} \\ \cmidrule(r){1-2}
					DS1 & Would you like to use more or less of your daily work for testing?						& SC   \\ 
					DS2 & Why do you want to spend more or less time for testing?       							& Free-text   \\ 
					DS3 & What kind of tests do you want to write/perform the most?         						& MC + free-text   \\ 
					DS4 & Would you spend more time on testing if it would be recognized?    & Likert 7 point   \\ 
					\bottomrule
				\end{tabular}%
			}
		\end{table}
		
		\subsection{Survey Tools}
		We implemented the survey using \textit{Prolific} to recruit participants and \textit{SoSci Survey} to host the survey itself.
		
		\subsubsection{Prolific} \label{sec:prolific}
		Prolific\footnote{\url{https://www.prolific.co/}} is an online platform to recruit participants for different kinds of studies like interviews and surveys. Together with Mechanical Turk (MTurk)\footnote{\url{https://www.mturk.com/}}, Prolific is one of the biggest recruitment platforms for participants. Prolific has some clear advantages to MTurk, since there are clear rules for both researchers and participants. All involved parties know about payments, obligations, and rights, and researchers also have better insights into the pool of possible participants~\cite{PALAN201822}. MTurk provides more participants (over 250,000), but most of them are located in the US~\cite{Robinson2019} while the more than 150,000 participants of Prolific are better distributed globally (\cref{sec:demo}). There is also empirical evidence that Prolific provides data of higher quality with less cheating and higher attention rates than MTurk~\cite{peer2022data}. 
		
		Recruiting participants with Prolific is not free, since both the participants and Prolific itself require payment. Our survey is set for an estimated completion time of ten minutes with an hourly rate of 10.50 \pounds, which means every respondent received 1.75~\pounds\ for completing the survey. The advantages of participants acquired by Prolific are that only preselected participants are permitted to take part in the survey and that they are motivated by their payment and approval score, which influences their future commissioning. Participants receive payment only after approval of their answers by the client. 
		
		\subsubsection{SoSci Survey}
		SoSci Survey\footnote{\url{https://www.soscisurvey.de}} is a powerful online platform to compose questionnaires with flexibility and individual design. SoSci Survey was designed for university research in 2003, has been under constant development ever since, and is free to use for researchers. The platform provides an easy, yet powerful editor for different kinds of questions and the collected data can be exported in various ways. 
		
		\subsection{Participants} \label{sec:participants}
		
		The target population of our survey consists of software
		developers since we want to understand their current and
		desired testing behavior.
		Prolific allows to pre-screen users based on demographic
		information as well as their self-declared expertise. We
		excluded users who do not work full- or part-time, and selected
		17 terms related to software engineering (e.g., debugging,
		version control) out of the hundreds of possible terms
		provided by Prolific to filter by relevant expertise.
		%
		%
		To increase trust in the participants' answers, we only
		accepted participants with an approval rate of 100\,\%. This
		rate is maintained by Prolific to keep track of how satisfied
		study conductors are with their respondents.

		After pre-screening, a pool of 9,156 eligible participants who
		had been active on Prolific in the last 90 days remained. Since we are
		only interested in (1) professional software developers who
		(2) currently work on a software project and (3) write code or
		tests, we used a pre-study (\cref{tab:allquestions}, IDs with PS) to
		filter the possible participants further. As users also
		receive payment for completing the pre-study, we requested 600
		responses to the pre-study from Prolific, of which 284
		answered all three pre-study questions in the affirmative. The
		final data is based on the responses of these 284 participants
		to the main survey, all of whom answered it completely.

		\subsection{Analysis of Responses}
		
		To analyze open-ended questions (PD6, CS5, DS2, AE1), we used \textit{qualitative content analysis}~\cite{DBLP:journals/libt/WhiteM06}. For each free-text question, we defined an empty set of codes/categories. While going through the answers manually, we added new categories whenever we encountered a new idea or perspective. If more than one participant mentioned the same idea, we used the same code for both (\cref{tab:codes}). Each code means that the participants mention this category in their answers. The coding was independently done by two researchers, who discussed and resolved all points of disagreement. 
		A summary of the analysis is given in Sections \ref{sec:projects} (Requirement for quality), \ref{sec:wellenough} (Tested well enough), and \ref{sec:morelesstime} (More or less time). The questions with ID \enquote{CS} (\cref{tab:allquestions}) are used to answer RQ1, while the one with ID \enquote{DS} is for RQ2.
		
		Closed-ended questions are analyzed by visualizing and bringing them into context with the research questions. We also measured the Spearman rank correlation matrix~\cite{kendall1948rank} as well as the multiple linear regression~\cite{freedman2009statistical} and ordinal logistic regression~\cite{mccullagh1980regression} matrices for all our variables to find dependent and significant variables.
		
		\begin{table}[]
			\caption{Codes found during the Qualitative content analysis}
			\label{tab:codes}
			\resizebox{\linewidth}{!}{%
				\begin{tabular}{llll}
					\toprule
					Requirement for quality (PD6) & Tested well enough (CS5) 			& More or less time (DS1) 				& Anything else (AE1)     \\ \midrule
					Customer Satisfaction	& Not valued 					& Skill		   						& Boring \\ 
					Mutation Score       	& No structure 					& Learn		   						& Business \\ 
					Review     				& Client 						& Like testing		   				& Data \\ 
					Bugs/Defects    		& Bugs 							& More important tasks		   		& Project \\ 
					Time    				& Data 							& Dislike testing		   			& Recognition \\ 
					Requirements    		& Resources 					& Process		   					& Skill \\ 
					Approval    			& Not needed 					& Satisfaction		   				& Testing \\ 
					Manual    				& More testing 					& Not enough testing				& Time \\ 
					Test Plan    			& - Missing parts	& Quality		   					&  \\ 
					Code Coverage    		& - Too big		& - Customer satisfaction	&  \\ 
					& - Too early		& - Solving issues  		&  \\ 
					& - Forgetting		& - Avoid issues  			&  \\ 
					& - Edge cases		& Right amount of testing  			&  \\ 
					& Skill 						&		   							&  \\ 
					& Time 							&		   							&  \\ 
				\end{tabular}%
			}
		\end{table}
		
		\subsection{Threats to Validity}
		Threats to \emph{internal validity} arise
		since the participants are distributed across several countries, the questions asked may be misunderstood or misinterpreted because of local differences in the language or because English is not their first language. This risk is reduced by our pilot study through which ambiguities in the questions were removed. The participants received the remuneration regardless of the time needed to finish the questionnaire, which may impact the quality of responses; however, respondents receive payment only if their answers are approved.
		Possible inconsistencies in the categorization of free-text responses were addressed by two researchers independently coding and resolving disagreements.
		
		There may also be threats to \emph{external validity}
		as the participants may not be distributed globally well enough to generalize the results, and the sampling algorithm of Prolific may be biased.
		Answers may not be from the perspective of a developer but from, e.g., a manager or consultant, since they may take different roles within their teams. However, since all participants stated they are currently coding or testing in a project, we assume they have insights of a developer.
		
		Threats to \emph{construct validity} arise from the design of the survey. The questions may not be specific enough to measure relationships between them and answer options may be missing. Despite the evaluation of the questionnaire in a pilot study, some questions and possible answers can nevertheless be misinterpreted by the participants because of missing explanations about tools and metrics asked. There may also be topics related to the research questions (e.g., whether the participants formal training or use test automation tools) that were not asked in the survey but could have given more insights into the subject, which is why an optional answer in the end was added (AE1).
		
		\section{Survey Results} \label{sec:results}
		
		\subsection{Demographics} \label{sec:demo}
		The survey questions provide demographic information at the level of individual participants as well as their projects.
		
		\subsubsection{Participants}
		The participants in our study exhibit a diverse range of ages, spanning from 18 to 52 years old. The majority of participants fall within the age range of 22 to 31, with a decreasing number of participants as age increases. A plausible explanation for this trend could be that as individuals progress in their software development careers, they may transition into managerial roles with reduced involvement in coding activities. This trend aligns with the Developer Survey conducted by Stack Overflow~\cite{stackoverflow}, where the majority of developers were between 18 and 34 years old, with few participants exceeding the age of 44.
		
		Approximately 31\,\% of our participants are students, which is expected considering a significant portion of our participants are under 30 years old. Interestingly, only 58\,\% of the students reported working part-time in a company, while the remaining 42\,\% work full-time
		while simultaneously pursuing their studies. 
		%
		The majority of our participants have already graduated with a university degree, with only 42 individuals reporting employment in a company with a High School diploma or equivalent (\cref{tab:userdemo}). With 80\,\% of the participants possessing a Bachelor's or Master's degree, the respondents slightly exceed the average qualification reported by the Developer Survey, where about 75\,\% reported a similar degree~\cite{stackoverflow}.
		
		The diverse range of experience levels from less than one year (9.5\,\%) to more than ten years (21.1\,\%) allows for a comprehensive exploration of different perspectives on testing. Additionally, representatives from various companies and company sizes (\cref{tab:userdemo}) providing insights into a wide array of development and testing processes.
		
		Current research~\cite{DBLP:conf/icse/BellerGZ15, DBLP:conf/issre/DakaF14, DBLP:conf/icse/MaximilienW03, aniche2022effective} and practice~\cite{nader2019executive, DBLP:conf/xpu/Bjerke-Gulstuen15} suggest that testing already needs to be done during the development phase, which implies that developers should at least perform unit testing to ensure the software's quality~\cite{beck2003test}. Consequently, we targeted developers specifically and not testers in this work, and indeed only 15 of our participants stated in their answers that there is a dedicated QA department rather than developer testing~\cite{tarlinder2016developer} in their company.
		While the majority of participants (62\,\%) identify their current role as developers, our dataset also includes consultants, testers, and managers. While these are not strictly software developers, we accepted them nevertheless with our pre-study criteria because of their active involvement in software development.

		\begin{table*}[]
			\caption{Demographics of the participants}
			\label{tab:userdemo}
			\resizebox{\linewidth}{!}{%
				\begin{tabular}{|lrr|lrr|}
					\hline
					\multicolumn{1}{|c|}{\textbf{Variable}}             & \multicolumn{1}{c|}{\textbf{\begin{tabular}[c]{@{}c@{}}Number of \\ participants\end{tabular}}} & \textbf{\begin{tabular}[c]{@{}c@{}}Number of participants \\ in percentage\end{tabular}} & \multicolumn{1}{c|}{\textbf{Variable}}                         & \multicolumn{1}{c|}{\textbf{\begin{tabular}[c]{@{}c@{}}Number of \\ participants\end{tabular}}} & \textbf{\begin{tabular}[c]{@{}c@{}}Number of participants \\ in percentage\end{tabular}} \\ \hline
					\multicolumn{1}{|l|}{\textbf{Country of Residence}} & \multicolumn{1}{c|}{}                                                                           &                                                                                          & \multicolumn{1}{l|}{\textbf{Highest type of graduation}}       & \multicolumn{1}{c|}{}                                                                           &                                                                                          \\ \hline
					Germany                                             & 6                                                                                               & 2.1\,\% \drawPercentageBar{0.021}                                                                                    & Ph.D.                                                          & 10                                                                                              & 3.5\,\% \drawPercentageBar{0.035}                                                                                    \\
					Hungary                                             & 9                                                                                               & 3.2\,\% \drawPercentageBar{0.032}                                                                                    & Master                                                         & 81                                                                                              & 28.5\,\% \drawPercentageBar{0.285}                                                                                    \\
					Greece                                              & 9                                                                                               & 3.2\,\% \drawPercentageBar{0.032}                                                                                    & Bachelor                                                       & 152                                                                                             & 53.5\,\% \drawPercentageBar{0.535}                                                                                    \\
					Netherlands                                         & 10                                                                                              & 3.5\,\% \drawPercentageBar{0.035}                                                                                    & High School                                                    & 42                                                                                              & 14.8\,\% \drawPercentageBar{0.148}                                                                                    \\ \cline{4-6} 
					Canada                                              & 10                                                                                              & 3.5\,\% \drawPercentageBar{0.035}                                                                                    & \multicolumn{1}{l|}{\textbf{Years of professional experience}} & \multicolumn{1}{c|}{}                                                                           &                                                                                          \\ \cline{4-6} 
					South Africa                                        & 12                                                                                              & 4.2\,\% \drawPercentageBar{0.042}                                                                                    & \textgreater 10                                                & 60                                                                                              & 21.1\,\% \drawPercentageBar{0.211}                                                                                    \\
					Mexico                                              & 12                                                                                              & 4.2\,\% \drawPercentageBar{0.042}                                                                                    & 6 - 10                                                         & 45                                                                                              & 15.8\,\% \drawPercentageBar{0.158}                                                                                    \\
					Spain                                               & 13                                                                                              & 4.6\,\% \drawPercentageBar{0.046}                                                                                    & 3 - 5                                                          & 89                                                                                              & 31.3\,\% \drawPercentageBar{0.313}                                                                                    \\
					Others                                              & 17                                                                                              & 6.0\,\% \drawPercentageBar{0.060}                                                                                    & 1 - 2                                                          & 64                                                                                              & 22.5\,\% \drawPercentageBar{0.225}                                                                                    \\
					Italy                                               & 21                                                                                              & 7.4\,\% \drawPercentageBar{0.074}                                                                                    & \textless 1                                                    & 27                                                                                              & 9.5\,\% \drawPercentageBar{0.095}                                                                                    \\ \cline{4-6} 
					United Kingdom                                      & 21                                                                                              & 7.4\,\% \drawPercentageBar{0.074}                                                                                    & \multicolumn{1}{l|}{\textbf{Employees in the company}}         & \multicolumn{1}{c|}{}                                                                           &                                                                                          \\ \cline{4-6} 
					Poland                                              & 27                                                                                              & 9.5\,\% \drawPercentageBar{0.095}                                                                                    & \textgreater 500                                               & 84                                                                                              & 29.6\,\% \drawPercentageBar{0.296}                                                                                    \\
					United States                                       & 36                                                                                              & 12.7\,\% \drawPercentageBar{0.127}                                                                                    & 251 - 500                                                      & 34                                                                                              & 12.0\,\% \drawPercentageBar{0.120}                                                                                    \\
					Portugal                                            & 74                                                                                              & 26.1\,\% \drawPercentageBar{0.261}                                                                                    & 101 - 250                                                      & 35                                                                                              & 12.3\,\% \drawPercentageBar{0.123}                                                                                    \\
					& \multicolumn{1}{l}{}                                                                            & \multicolumn{1}{l|}{}                                                                    & 51 - 100                                                       & 24                                                                                              & 8.5\,\% \drawPercentageBar{0.085}                                                                                    \\
					& \multicolumn{1}{l}{}                                                                            & \multicolumn{1}{l|}{}                                                                    & 26 - 50                                                        & 24                                                                                              & 8.5\,\% \drawPercentageBar{0.085}                                                                                    \\
					& \multicolumn{1}{l}{}                                                                            & \multicolumn{1}{l|}{}                                                                    & 11 - 25                                                        & 30                                                                                              & 10.6\,\% \drawPercentageBar{0.106}                                                                                    \\
					& \multicolumn{1}{l}{}                                                                            & \multicolumn{1}{l|}{}                                                                    & \textless 10                                                   & 54                                                                                              & 19.0\,\% \drawPercentageBar{0.190}                                                                                    \\ \hline
				\end{tabular}
			}
		\end{table*}
		
		The participants come from diverse geographical locations worldwide, with a particular emphasis on Europe and North America. The distribution of participants is detailed in \cref{tab:userdemo}, which also includes additional demographic information. Interestingly, over 25\,\% of the participants are from Portugal,
		but the study encompasses participants from countries outside of Europe, such as South Africa, New Zealand, and Chile, contributing to a globally representative sample.
		
		Our study includes data from 20\,\% female participants. While this figure falls below the average of women in computer science with 25\,\% reported previously~\cite{murphy2019examining}, the same study found a general decline in female computer science graduates by 8\,\% to 17\,\%. This suggests that our study achieves an average or potentially higher representation of female participants following the reported decline~\cite{murphy2019examining}.

		\subsubsection{Projects} \label{sec:projects}
		\begin{table*}[]
			\caption{Demographics of the projects}
			\label{tab:projectdemo}
			\resizebox{\linewidth}{!}{%
				\begin{tabular}{|lrr|lrr|}
					\hline
					\multicolumn{1}{|c|}{\textbf{Variable}}             & \multicolumn{1}{c|}{\textbf{\begin{tabular}[c]{@{}c@{}}Number of \\ participants\end{tabular}}} & \multicolumn{1}{c|}{\textbf{\begin{tabular}[c]{@{}c@{}}Number of participants \\ in percentage\end{tabular}}} & \multicolumn{1}{c|}{\textbf{Variable}}                 & \multicolumn{1}{c|}{\textbf{\begin{tabular}[c]{@{}c@{}}Number of \\ participants\end{tabular}}} & \textbf{\begin{tabular}[c]{@{}c@{}}Number of participants \\ in percentage\end{tabular}} \\ \hline
					\multicolumn{1}{|l|}{\textbf{Domain}}               & \multicolumn{1}{l|}{}                                                                           &                                                                                                               & \multicolumn{1}{l|}{\textbf{Employees in the project}} & \multicolumn{1}{l|}{}                                                                           & \multicolumn{1}{l|}{}                                                                    \\ \hline
					Military                                            & 2                                                                                               & 0.7\,\% \drawPercentageBar{0.007}                                                                                                         & \textless 5                                            & 106                                                                                             & 37.3\,\% \drawPercentageBar{0.373}                                                                                    \\
					Insurance                                           & 5                                                                                               & 1.8\,\% \drawPercentageBar{0.018}                                                                                                         & 5 - 9                                                  & 89                                                                                              & 31.3\,\% \drawPercentageBar{0.313}                                                                                    \\
					Automotive                                          & 13                                                                                              & 4.6\,\% \drawPercentageBar{0.046}                                                                                                         & 10 - 19                                                & 48                                                                                              & 16.9\,\% \drawPercentageBar{0.169}                                                                                    \\
					Food                                                & 14                                                                                              & 4.9\,\% \drawPercentageBar{0.049}                                                                                                         & 20 - 50                                                & 25                                                                                              & 8.8\,\% \drawPercentageBar{0.088}                                                                                    \\
					Government                                          & 15                                                                                              & 5.3\,\% \drawPercentageBar{0.053}                                                                                                         & \textgreater 50                                        & 17                                                                                              & 6.0\,\% \drawPercentageBar{0.060}                                                                                    \\ \cline{4-6} 
					Entertainment                                       & 16                                                                                              & 5.6\,\% \drawPercentageBar{0.056}                                                                                                         & \multicolumn{1}{l|}{\textbf{Team sizes}}               & \multicolumn{1}{l|}{}                                                                           & \multicolumn{1}{l|}{}                                                                    \\ \cline{4-6} 
					Health                                              & 19                                                                                              & 6.7\,\% \drawPercentageBar{0.067}                                                                                                         & 1                                                      & 11                                                                                              & 3.9\,\% \drawPercentageBar{0.039}                                                                                    \\
					Logistics                                           & 21                                                                                              & 7.4\,\% \drawPercentageBar{0.074}                                                                                                         & 2 - 3                                                  & 76                                                                                              & 26.8\,\% \drawPercentageBar{0.268}                                                                                    \\
					Games                                               & 22                                                                                              & 7.7\,\% \drawPercentageBar{0.077}                                                                                                         & 4 - 6                                                  & 109                                                                                             & 38.4\,\% \drawPercentageBar{0.384}                                                                                    \\
					Public Services                                     & 24                                                                                              & 8.5\,\% \drawPercentageBar{0.085}                                                                                                         & 7 - 10                                                 & 68                                                                                              & 23.9\,\% \drawPercentageBar{0.239}                                                                                    \\
					Finance                                             & 38                                                                                              & 13.4\,\% \drawPercentageBar{0.134}                                                                                                         & \textgreater 10                                        & 21                                                                                              & 7.4\,\% \drawPercentageBar{0.074}                                                                                    \\ \cline{4-6} 
					Communication                                       & 40                                                                                              & 14.1\,\% \drawPercentageBar{0.141}                                                                                                         & \multicolumn{1}{l|}{\textbf{Code Coverage}}            & \multicolumn{1}{l|}{}                                                                           & \multicolumn{1}{l|}{}                                                                    \\ \cline{4-6} 
					Other                                               & 56                                                                                              & 19.7\,\% \drawPercentageBar{0.197}                                                                                                         & Not measured/available                                 & \multicolumn{1}{r}{90}                                                                          & \multicolumn{1}{r|}{31.7\,\% \drawPercentageBar{0.317}}                                                               \\ \cline{1-3}
					\multicolumn{1}{|l|}{\textbf{Quality measurements}} & \multicolumn{1}{l|}{}                                                                           &                                                                                                               & 0\% - 20\%                                             & \multicolumn{1}{r}{19}                                                                          & \multicolumn{1}{r|}{6.7\,\% \drawPercentageBar{0.067}}                                                               \\ \cline{1-3}
					Code Coverage                                       & \multicolumn{1}{r}{130}                                                                         & \multicolumn{1}{r|}{45.8\,\% \drawPercentageBar{0.458}}                                                                                    & 21\% - 40\%                                            & \multicolumn{1}{r}{23}                                                                          & \multicolumn{1}{r|}{8.1\,\% \drawPercentageBar{0.081}}                                                               \\
					Mutation Score                                      & \multicolumn{1}{r}{23}                                                                          & \multicolumn{1}{r|}{8.1\,\% \drawPercentageBar{0.081}}                                                                                    & 41\% - 60\%                                            & \multicolumn{1}{r}{49}                                                                          & \multicolumn{1}{r|}{17.3\,\% \drawPercentageBar{0.173}}                                                               \\
					Defects opened/closed                               & \multicolumn{1}{r}{156}                                                                         & \multicolumn{1}{r|}{54.9\,\% \drawPercentageBar{0.549}}                                                                                    & 61\% - 80\%                                            & \multicolumn{1}{r}{71}                                                                          & \multicolumn{1}{r|}{25.0\,\% \drawPercentageBar{0.250}}                                                               \\
					None                                                & \multicolumn{1}{r}{60}                                                                          & \multicolumn{1}{r|}{21.1\,\% \drawPercentageBar{0.211}}                                                                                    & 81\% - 100\%                                           & \multicolumn{1}{r}{33}                                                                          & \multicolumn{1}{r|}{11.6\,\% \drawPercentageBar{0.116}}                                                               \\
					Other                                               & \multicolumn{1}{r}{14}                                                                          & \multicolumn{1}{r|}{4.9\,\% \drawPercentageBar{0.049}}                                                                                    &                                                        & \multicolumn{1}{l}{}                                                                            & \multicolumn{1}{l|}{}                                                                    \\ \hline
				\end{tabular}
			}
		\end{table*}
		
		The global distribution of participants results in a wide variety of software projects spanning different domains. \Cref{tab:projectdemo} provides an overview of project demographics and highlights that the majority of projects are concentrated in the communication and finance domains. However, numerous other domains, such as education, robotics, fashion, and real estate, are also represented in the dataset.
		A significant proportion of projects (70\,\%) are realized by teams with fewer than ten members, while only 6\,\% of the projects involve more than 50 staff members. This suggests that many projects might have adopted agile methodologies using small teams~\cite{schwaber2002agile}.
		
		\subsection{RQ 1: How do developers engage with testing in practice?}
		
		\begin{figure}[t]
			\includegraphics[width=\linewidth]{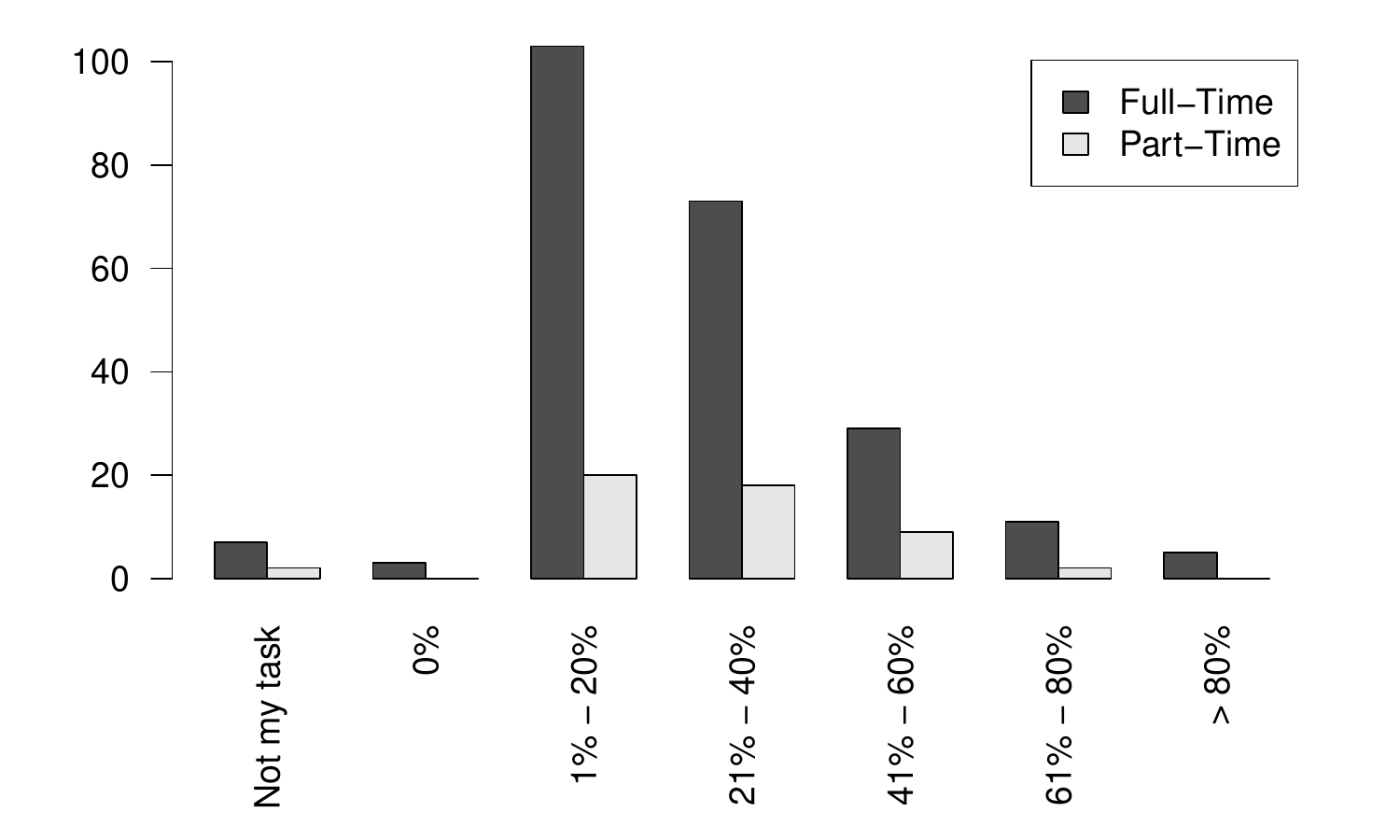}
			\centering
			\caption{Proportion of the working time that is used for testing with full- and part-time jobs by the participants (CS1)}
			\label{fig:testingtime}
		\end{figure}
		
		\emph{1) How much time do developers invest in testing?}
		The time invested in testing activities among participants (CS1) reveals that a majority (more than 80\,\%) dedicate relatively limited time to testing (less than 40\,\%) during their work hours (\cref{fig:testingtime}). Additionally, a great portion (more than 44\,\%) allocates even less time to testing (less than 20\,\%), which will be further analyzed and investigated in \cref{sec:morelesstime}.
		
		\begin{table}[]
			\caption{Kind of tests the participants are performing versus what they want to perform} 
			\label{tab:testkind}
			\resizebox{\linewidth}{!}{%
				\begin{tabular}{|lrrrr|}
					\hline
					\multicolumn{1}{|l|}{\multirow{2}{*}{Kind of tests}} & \multicolumn{2}{c|}{Tests currently performed}                & \multicolumn{2}{c|}{Tests wanted to perform} \\ \cline{2-5} 
					\multicolumn{1}{|l|}{}                               & \multicolumn{1}{c|}{Number} & \multicolumn{1}{c|}{Percentage} & \multicolumn{1}{c|}{Number}   & Percentage   \\ \hline
					Unit tests                                           & 147                         & 51.8\,\% \drawPercentageBar{0.518}                           & 139                           & 48.9\,\% \drawPercentageBar{0.489}        \\
					Integration tests                                    & 121                         & 42.6\,\% \drawPercentageBar{0.426}                           & 81                            & 28.5\,\% \drawPercentageBar{0.285}        \\
					System tests                                         & 98                          & 34.5\,\% \drawPercentageBar{0.345}                           & 71                            & 25.0\,\% \drawPercentageBar{0.250}        \\
					API tests                                            & 91                          & 32.0\,\% \drawPercentageBar{0.320}                           & 64                            & 22.5\,\% \drawPercentageBar{0.225}        \\
					UI tests                                             & 102                         & 35.9\,\% \drawPercentageBar{0.359}                           & 66                            & 23.2\,\% \drawPercentageBar{0.232}        \\
					Manual tests                                         & 158                         & 55.6\,\% \drawPercentageBar{0.556}                           & 68                            & 23.9\,\% \drawPercentageBar{0.239}        \\
					Smoke tests                                          & 27                          & 9.5\,\% \drawPercentageBar{0.095}                           & 17                            & 6.0\,\% \drawPercentageBar{0.060}        \\
					Sanity tests                                         & 28                          & 9.9\,\% \drawPercentageBar{0.099}                           & 21                            & 7.4\,\% \drawPercentageBar{0.074}        \\
					Regression tests                                     & 67                          & 23.6\,\% \drawPercentageBar{0.236}                           & 46                            & 16.2\,\% \drawPercentageBar{0.162}        \\
					Other                                                & 7                           & 2.5\,\% \drawPercentageBar{0.025}                           & 12                            & 4.2\,\% \drawPercentageBar{0.042}        \\ \hline
				\end{tabular}
			}
		\end{table}
		
		\emph{2) What kind of tests do developers write?}
		The respondents apply a variety of testing approaches (CS2). \Cref{tab:testkind} illustrates that both unit and manual testing are carried out by over half of the participants (51.8\,\%), making them the primary test types followed by integration tests with 42.6\,\%. Other types of tests, such as smoke or sanity tests, are less commonly applied, each accounting for less than 10\,\%. The reasons for this disparity could be multifaceted, such as a lack of training or the perception that functional testing is more important.
		
		\begin{figure}[t]
			\includegraphics[width=\linewidth]{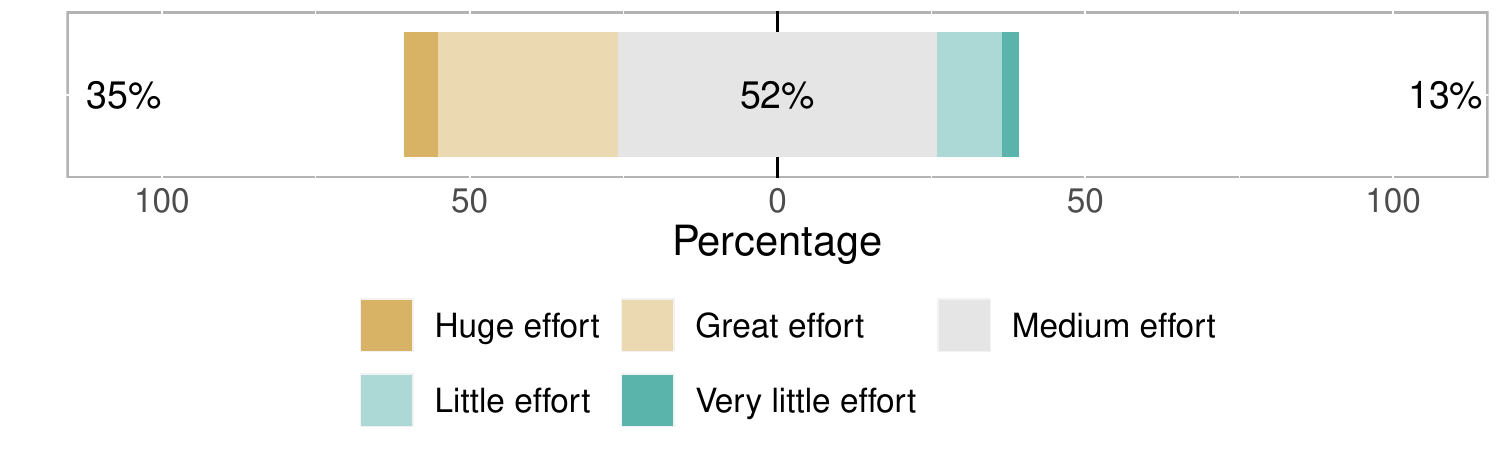}
			\centering
			\caption{Effort needed for testing (CS6)}
			\label{fig:effort}
		\end{figure}
		
		\begin{figure}[t]
			\includegraphics[width=\linewidth]{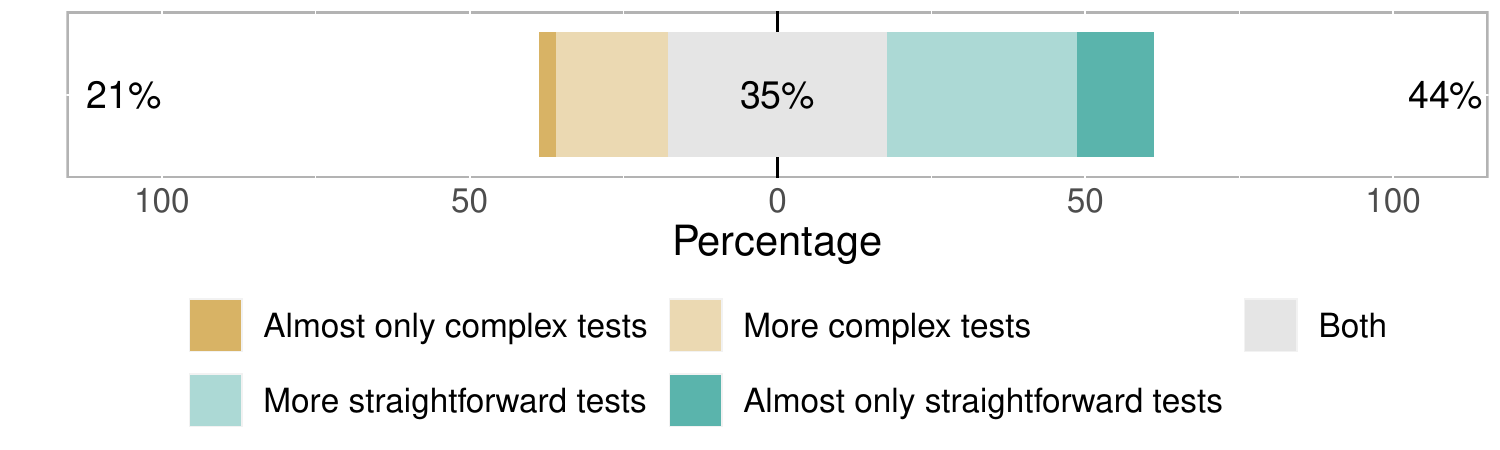}
			\centering
			\caption{Complexity of the tests written by participants (CS7)}
			\label{fig:complexity}
		\end{figure}
		
		\begin{figure}[t]
			\includegraphics[width=\linewidth]{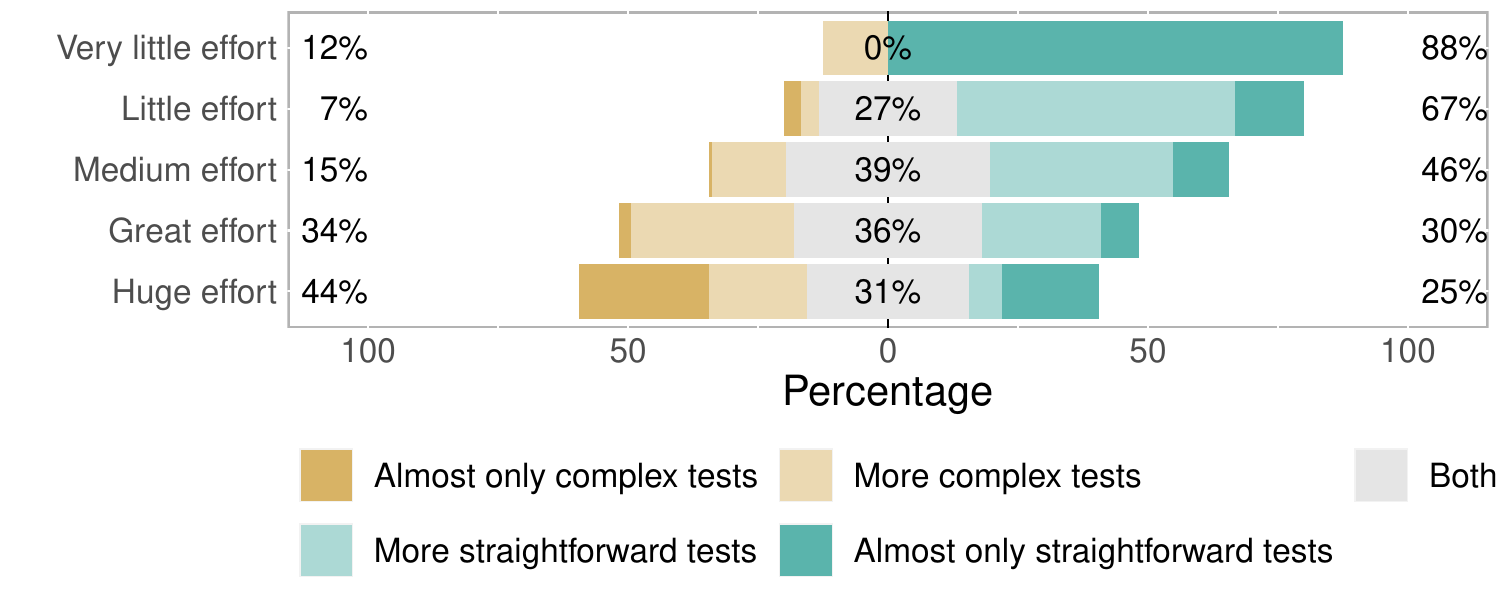}
			\centering
			\caption{Complexity and effort combined from questions CS6 and CS7}
			\label{fig:complexity_effort}
		\end{figure}
		
		\emph{3) How much effort does it take to write tests?}
		The level of effort required for writing tests in participants' projects is illustrated in \cref{fig:effort} (CS6). The data reveals that only 13\,\% of participants consider testing to be easy while more than half find testing to be challenging, indicating moderate effort involved. On the other hand, 35\,\% face significant issues in testing, as it demands great effort to test their project.
		
		The complexity of tests is influenced by the effort required (CS7). Among the participants, 44\,\% primarily write simpler tests, maybe aiming to achieve quality goals with minimal effort (\cref{fig:complexity}). Roughly one-third write both complex and simple tests, while only 21\,\% develop complex tests to ensure quality and address edge cases (based on answers of CS5).
		
		\Cref{fig:complexity_effort} highlights the relationship between the complexity of tests and the effort expended to write them,  indicating that as the complexity of tests increases, the effort to write them increases, too. An intuitive explanation would be that poor design amplifies the effort required for writing tests, resulting in more intricate tests and discouraged developers. 
		
		\emph{4) Do developers follow certain quality metrics and requirements?}
		Code coverage
		is a widely utilized metric for quality~\cite{DBLP:conf/qrs/Hemmati15} and a first indicator of a project's testing state. Nevertheless, only 45\,\% of the participants' projects use code coverage (PD4). In certain cases, answers indicate that measuring coverage may be infeasible, such as with embedded software, or too computationally expensive~\cite{ivankovic2019coverage}.
		Overall, developers may lack awareness of the true quality of their projects due to insufficient or inadequate metrics, which we believe can influence their motivation to test.
		
		Furthermore, only 27\,\% of participants' projects have specific quality requirements or goals, as indicated by question PD6 (\cref{tab:allquestions}): 31 participants mention code coverage as their quality goal, while others emphasize the need for minimal defects or the basic functionality of the software, which may not be easily measurable. Some participants rely on management approval or feedback from quality assurance departments, while others prioritize customer satisfaction or peer reviews. Two projects even use mutation scores as their testing goal.
		Different forms of reviews, evaluations by superiors, or input from customers, are used, too. In certain projects, timely completion plays a crucial role to meet deadlines, even at the expense of quality. Projects that undergo requirements engineering in the first place rely on the fulfillment of those or adherence to a test plan as their indicators of quality. Moreover, the lack of specific quality goals may result in misguided assumptions about code quality and the overall necessity of testing since their perception of quality is based on their personal feeling rather than a quality metric.

		\emph{5) Do developers believe their projects are sufficiently tested?} \label{sec:wellenough}
		Approximately 60\,\% of the developers expressed confidence in the level of testing conducted in their projects (CS5). This percentage is notably high considering the limited amount of time dedicated to testing by the participants.
		
		However, many developers also highlighted various factors that contribute to perceived inadequacies in testing.
		One common concern raised by developers in the free-text field of CS5 is the undervaluation of testing within their companies. Developers often assign higher priority to other tasks and allocate their time accordingly, neglecting comprehensive testing. Another factor mentioned to influence the perceived lack of testing is the absence of well-defined test processes, which is thought to lead to issues such as poor documentation, insufficient communication between developers and testers, and overall inadequate testing practices.
		
		Nominal client involvement was cited as a reason for minimal engagement with testing or a focus on new features because there is no requirement to write tests. Additionally, some developers believe their software lacks proper testing because evidence shows that the software still contains bugs, unresolved defects, or uses artificial data for testing. Projects with untested components or pending evaluation of critical edge cases are also cited as requiring more testing.
		
		The lack of time and resources emerges from CS5 as a recurring constraint that hampers proper testing. Many participants express a desire to engage in testing activities but are constrained by factors beyond their control.
		Overall, developers' perceptions of sufficient testing vary and are influenced by factors such as prioritization, development process maturity, client engagement, and resource limitations.
		
		\begin{figure}[t]
			\includegraphics[width=\linewidth]{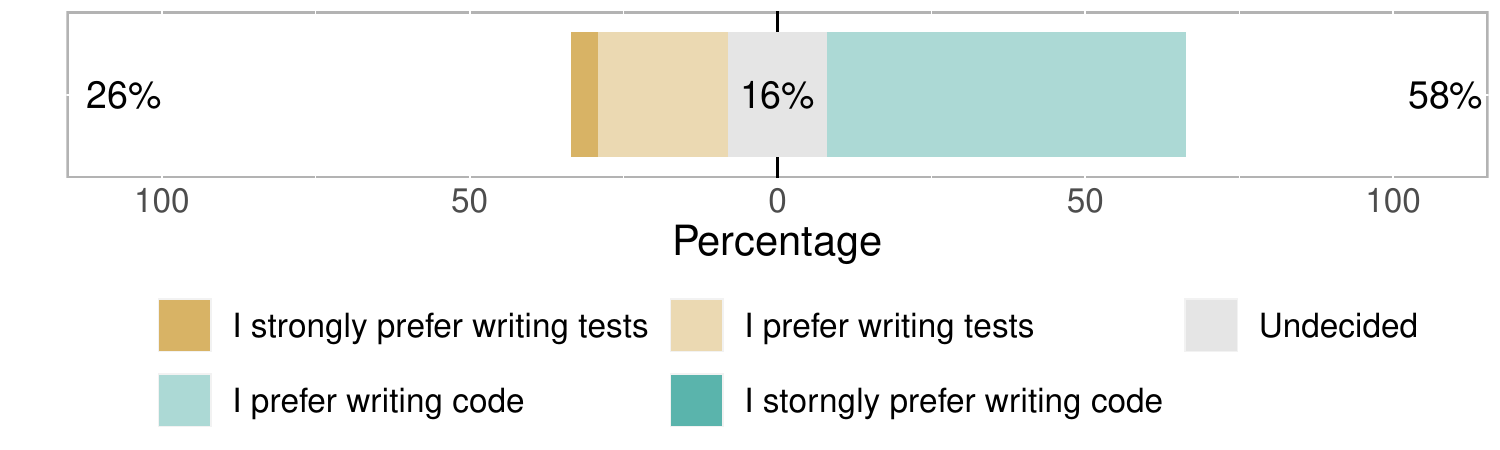}
			\centering
			\caption{The preference to write code or tests (CS3)}
			\label{fig:writetestcode}
		\end{figure}
		
		\begin{figure}[t]
			\includegraphics[width=\linewidth]{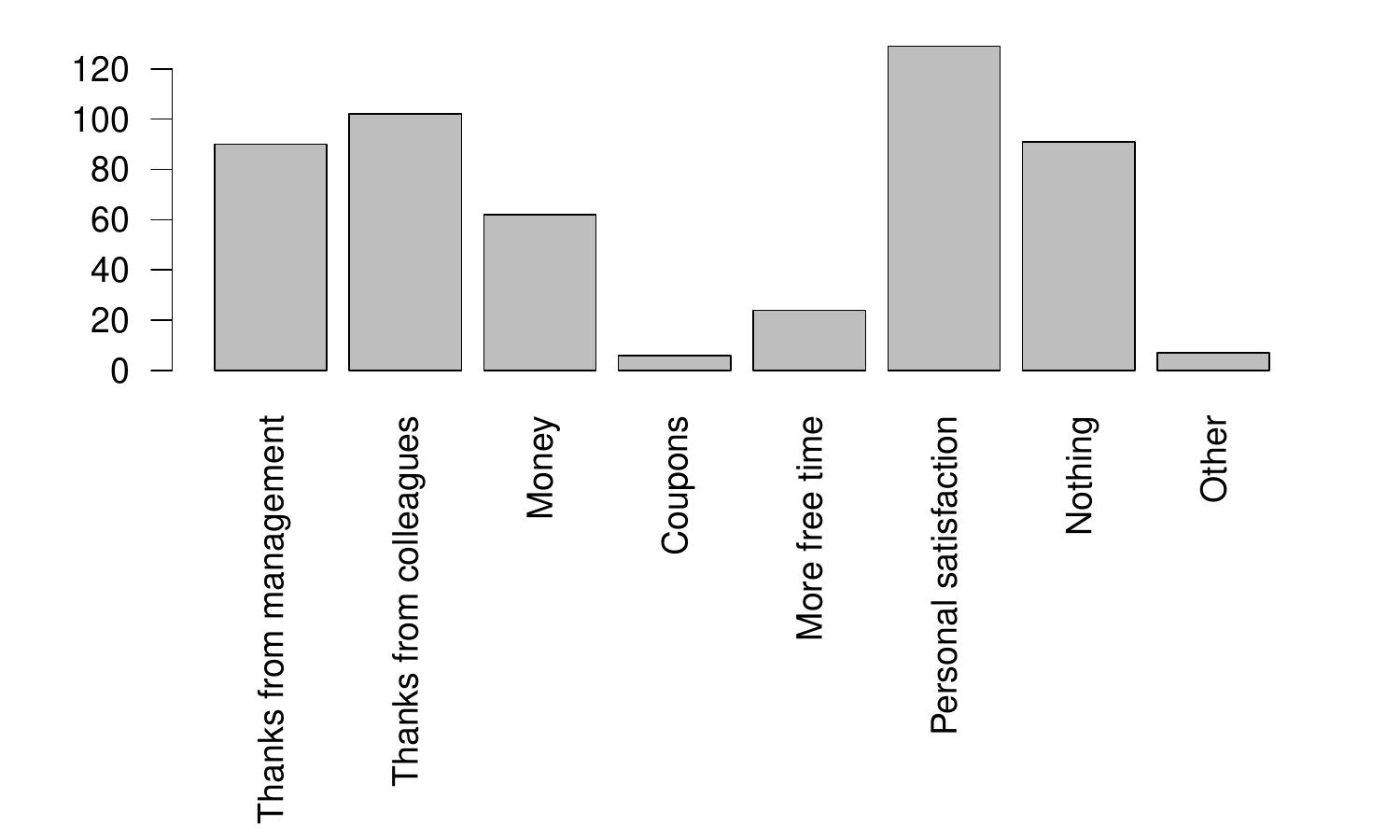}
			\centering
			\caption{Recognition for testing (CS4)}
			\label{fig:recognition}
		\end{figure}
		
		\emph{6) How is testing recognized?}
		Approximately one-third of developers receive an acknowledgment from both management and colleagues (\cref{fig:recognition}), demonstrating that testing is valued in their respective companies and projects while notable 25\,\% of developers receive monetary rewards for their testing efforts in addition (CS4). Furthermore, different types of external recognition, or the lack thereof, are stated by the participants in question AE1. One participant stated that more code coverage in the project proved to lead to fewer bugs in the remaining project and that a \enquote{reputation of writing 'bug-free' software (or as close as possible)} (P65) was given to the participant for it. Enough testing does not only improve quality but also shows that the software is working as intended efficiently and the risk of failing the project is minimized (P193). Some describe that their company understands the importance of software testing (e.g., P33), but there are many more who do not (e.g., P95).
		
		Nearly half of the participants (around 50\,\%) claim to be intrinsically motivated and test for personal satisfaction, even though only a minority of the developers see testing as a crucial part of the development process and are engaged in testing and enjoy it. For example, one participant compared testing with \enquote{a fun puzzle to figure out} (P65) and another is satisfied with the little effort required for testing because their software is designed for it (P11). Discovering hidden bugs and issues that would have gone into production without sufficient testing are also mentioned as motivating (P66), as well as the resulting time savings (P48). Thoroughly testing the participant's software gives the developers peace of mind when they are sure \enquote{they did not mess things up} (P61). A good test suite also improves the ability to refactor the code base when the application slowly evolves over the years, because bugs introduced during maintenance can be found by the existing test suite (P65). It also gives the developers personal satisfaction when their code works smoothly and users can work without frustration (P65). We believe that this intrinsic motivation serves as a driving force for writing tests and investing effort beyond what is formally required.
		
		On the other hand, it is concerning that approximately one-third of developers do not receive any form of external recognition or have no intrinsic motivation for testing (e.g., P95). This lack of acknowledgment and motivation can potentially result in a reduced inclination to write tests, ultimately impacting the overall software quality. In addition, the majority of developers (58\,\%, \cref{fig:writetestcode} prefer writing code over tests (CS3).
		
		We applied both correlation and regression analysis to the demographic variables about the participants (\cref{tab:allquestions}, IDs with UD) and the projects (IDs with PD), and the variables of the current state of testing in their project (IDs with CS). Unfortunately, no significant or dependent variables could be found to explain our findings. Consequently, we cannot report whether the current state of testing is dependent on any of the participant or project demographics we considered.

		\summary{RQ 1}{Testing is not the favorite task of developers. The time invested in testing is limited and sometimes takes great effort. Even though many believe their projects are tested well enough, common demotivating factors are higher prioritized tasks, bad communication, and missing recognition.}
		
		\subsection{RQ 2: How would developers actually like to engage with testing?} \label{sec:morelesstime}
		
\begin{figure}[t]
			\includegraphics[width=\linewidth]{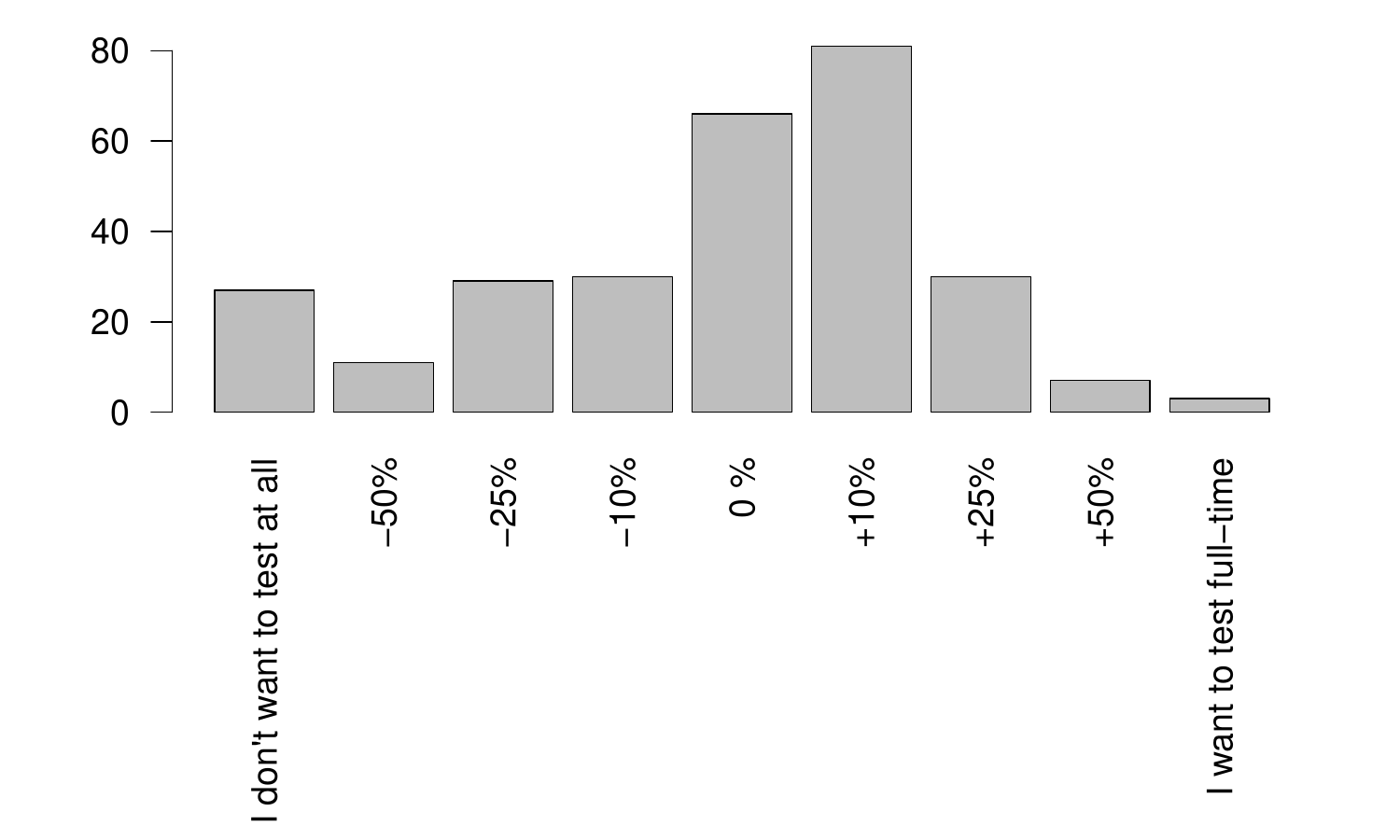}
			\centering
			\caption{Additional testing time the participants want to use (DS1)}
			\label{fig:additionaltestingtime}
		\end{figure}

		The participants are split (DS1) between those who want to test more, those who do not, and those who are satisfied with their current testing efforts (\cref{fig:additionaltestingtime}). While 42\,\% of the participants want to test more, there are 58\,\% who do not, either because they want to test less (24\,\%) or because they are satisfied with their current testing effort (34\,\%).
		
		\emph{1) Why do developers want to test more?}
		Most of the 42\,\% who want to test more estimate they would use 10\,\% more of their time for testing.
		The more time participants already use for testing, the less they want to reduce it. Most of those respondents not testing at all do not want to start with it. In contrast, the more time the participants already invest in testing, the less they want to reduce this amount. This could mean the more they test the more they understand that testing increases the quality and decreases their workload later.

                \begin{figure}[t]
			\includegraphics[width=\linewidth]{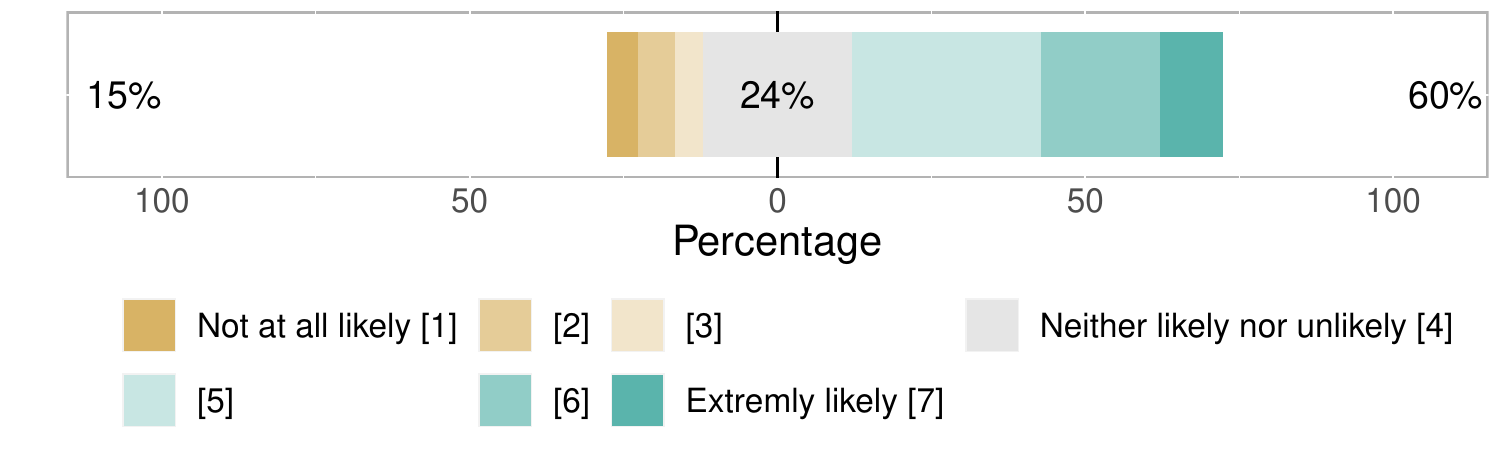}
			\centering
			\caption{More or less testing if better recognized (DS4)}
			\label{fig:morerecognition}
		\end{figure}

		To investigate why developers want to test more in more detail, we analyze the free-text answers to the mandatory question of why they want to test more or less (DS2), where we observe the following reasons:
		\begin{itemize}
			\item Find  more issues, blockers, and edge cases (37)
			\item Ensure and improve quality  (35)
			\item Like to test their own and their colleagues' code (17)
			\item Decreased maintenance expenses (15)
			\item Satisfied to show others that their code is working fine and better than other's (9)
			\item Improve their coding and testing skills (5)
			\item Try new aspects of development and new technologies~(1)
			\item Improve strategic thinking and solve problems (1)
			\item Increase variety in their daily work (1)
			\item Find areas for improvements and refactoring (1)
			\item Incorporate new team members faster (1)
		\end{itemize}
		The observation that lack of recognition inhibits motivation is confirmed by question (DS4), which shows that nearly 60\,\% of the participants would test more if it were better recognized (\cref{fig:morerecognition}). Overall, the environment seems to have a greater impact on developers than their personal motivation.
		
		\suggestion{Suggestion 1}{The recognition of testing within companies should be increased for both developers and at the management level to motivate more engagement with testing.}
		
		The answers to question AE1 provide further insights into why developers have a desire to test more: They appreciate the positive outcomes of testing, such as improved quality (P239), as well as the opportunity to solve bugs (P67). In one particular company, the quality assurance team is larger than the development team (P17), which leads to management valuing the time, resources, and effort dedicated to testing (P17). Many participants are motivated to test because they have access to a testing environment that closely resembles the production environment (P11), there is effective communication between developers and testers (P215), and they receive feedback from end-users (P268). Some participants have already completed software testing courses provided by their company and see the benefits of their involvement in the project they are working on (P45). These participants express a need and desire for training in writing software tests and managing test suites and infrastructure, as they recognize the potential benefits (e.g., P45). One participant describes that they have a very good policy for testing, where everyone gets trained to write tests (P33). Some participants see testing as an integral task inside the development cycle (P27) where all code should be accompanied by unit tests (P194), part of a review system (P194), and all tests should run continuously (P218).
		
		\emph{2) Why are developers satisfied with their current testing efforts?}
		Almost one-quarter of the participants are satisfied with the amount of time currently invested in testing (\cref{fig:additionaltestingtime}). Of these 66 developers, only eight do not write tests and also do not want to.
		Participants mention a well-set-up and well-handled testing environment or that they are already in maintenance without developing new features. Three participants also point out that there is a dedicated tester in their team or even a separate quality assurance department for testing. There are also projects which currently meet the given quality goal or the developers can choose their used time for testing. Those answers are reflected in the free-text answers to DS2:
		\begin{itemize}
			\item Amount of testing is adequate (33)
			\item Another person/department responsible for testing (3)
			\item Quality goal reached (2)
			\item Well set-up testing process (1)
			\item Project in maintenance (1)
		\end{itemize}
		
		\emph{3) Why do developers want to test less?}
		Out of all respondents, 34\,\% want to test less than currently, with almost the same number in the categories -10\,\%, -25\,\%, and participants who do not want to test at all (DS1). Especially the percentage of participants who do not want to test at all with almost 10\,\% is high compared to the participants who want to test full-time at about 1\,\%, which means there is a great contrast between engagement to write tests and code. Many developers do not mind testing in general but do not want to test the code of their colleagues, because it is not well-written or hard to understand. 
		
		\suggestion{Suggestion 2}{Developers should always write unit tests for their own code to support other developers working with the code, who are unlikely to write these tests.}
		
		In comparison with the currently performed tests in the participants' projects (\cref{tab:testkind}), the number of tests the participants want to perform decreases (DS3). Unit testing is least affected (3\,\%)  which confirms that unit tests are the most popular kind of tests performed by developers, while manual testing is the least popular one decreasing more than 50\,\%.
		
		In the free-text responses, we identified several reasons why developers want to engage less with testing (DS2):
		\begin{itemize}
			\item Like coding better and dislike testing (70)
			\item No working testing process and infrastructure (30)
			\item Do more important things and  meet deadlines (21)
			\item Testing as boring, frustrating, and repetitive activity (19)
			\item Lack of testing skills (15)
			\item Missing resources to test well enough (13)
			\item No proper training in testing (4)
			\item Project is tested well enough (1)
			\item Project is too small or old to be tested (1)
			\item Software will be replaced soon (1)
			\item There are other developers in charge of testing (1)
			\item Find manual testing exhausting and unnecessary (1)
			\item Lack of communication (1)
			
		\end{itemize}
		
		During the analysis of the answers to question AE1, we discovered more detailed insights into why developers want to test less. For example, manual tests are considered boring and time-wasting by the respondents (e.g., P49), especially when they have to be done over and over again when versions change (P27). Since many participants consider testing as boring, they think \enquote{it takes time away from more important issues} (P125) like implementing new features (P137). Because the participants do not want to test these new features either, more bugs may be introduced (P218). The development process in general is depicted as not thought through (P68) which causes the developers to de-prioritize testing (P140) and dislike the current process (P121). Another demotivating factor is a lack of infrastructure like continuous integration (P49), and that tests are badly implemented by colleagues and predecessors (P196), which causes disengagement from testing.
		
		\suggestion{Suggestion 3}{Whenever feasible, software tests should be designed to be automatically executed and supported by a continuous integration infrastructure.}
		
		Furthermore, managers are mentioned to be interested in testing only when \enquote{a major or critical bug that is disruptive to the business is found} (P12), only to get back to old habits when the crisis is averted. Everything the customer does not recognize or value with money will be neglected or de-prioritized (P252). Especially in start-ups or small companies, each employee has more than one task in the company (P86), which can cause the testing to be forgotten (P86). Others have to use \enquote{in-house developed tools} (P104) which have low usability or simply do not work as needed. This prevents developers from engaging with testing as they want to (P90).
		
		\suggestion{Suggestion 4}{Awareness at the management level needs to be raised about the importance of testing activities to increase and maintain the quality of software.}
		
		Lack of test data and scenarios is mentioned as a source of problems, for example when there is insufficient test data that is as close as possible to production data (P164). Since many of the participants only use fake, sample, or mocked data, their tests are not considered \enquote{completely trustworthy} (P4) to them. The testers also need a variance of test data to include edge cases in their test suites, but most of the projects and customers only provide generic test data without covering special cases (P164). Since the developers do not know the interactions of the end-users with the application, either because the product is not released yet (P42) or because user data is not collected (P73), they cannot create test scenarios that are as close as possible to reality. The lack of these scenarios can lead to users breaking the system even if it has been tested well, and developers then receive the blame from management (P42). In addition, the lack of important information and skills (P26) makes it hard to write good and sustainable test suites (P26), and developers get frustrated by all the failures and changes because of missing data and communication (P138).
		
		\suggestion{Suggestion 5}{Requirements, common scenarios and problems should be communicated among all stakeholders, including developers, to decrease the time for troubleshooting.} 
		
		Software in some domains seems particularly difficult to test, like games (P74), simulators (P74), embedded devices (P200), or software with hardware requirements (in particular when the required hardware is not available, P128). There are also old projects that contain many bugs and lack a testing infrastructure (P78), which are mostly replaced with new ones including an automated test suite and documentation (P78).
		
		\suggestion{Suggestion 6}{Developers require sufficient test data and robust infrastructure for automated test executions to write tests and avoid demotivational factors.}
		
		In addition to identifying various blockers, we also observed different attitudes among the developers. For instance, one participant believes that others who work fewer hours should be responsible for writing tests (\enquote{There are people working less than me so they should be doing the testing}, P69). Some participants consider their projects too small to warrant testing at all(P3). On the other hand, one of the participants stated to prioritize simplicity and reliability in their code rather than creating a complex and robust product (P121).
		
		\suggestion{Suggestion 7}{Developers should be made aware of the significance of testing, ensuring they understand the advantages of testing as well as the drawbacks of neglecting it.}
		
		We also noticed that many developers are frustrated by the lack of time (P140) and training (P38) they are given. Not only are lack of testing skills in general mentioned, but some think it is especially hard to write tests for edge cases (P185). Some participants would like to have more or better training in writing good tests (e.g., P38) which may result in better test suites and an increase in the quality of their products. Another important issue is the lack of time for writing tests given by the management or the customer (P65). Some participants consider this as the main problem when the quality of the software does not meet the excepted one (P44). Others have enough time for testing and a test suite, but it takes too long to execute all of them (P187).
		
		\suggestion{Suggestion 8}{Developers would benefit from software testing training programs to gain the necessary skills, knowledge and mindset to execute effective testing practices.}
		
		We applied both correlation and regression analysis to the demographic variables about the participants (\cref{tab:allquestions}, IDs with UD) and the projects (IDs with PD), and the variables of the current (IDs with CS) and desired state of testing in their project (IDs with DS). Unfortunately, also in this case no significant or dependent variables could be found to explain our findings. This means that we cannot make any statements about how the desired state of testing depends on the current state of testing or any of the participant or project demographics we asked for.
		
		\summary{RQ 2}{Developers claim they would write more tests to ensure quality and increase personal satisfaction, but would like to receive better recognition for this. On the other hand, many developers want to test less than they currently do since they perceive testing as boring compared to other tasks.}
		
		\section{Related Work}
		
		Most of the time developers spend in their IDEs, they are reading,
		writing, and modifying their code. Only about 9\,\% of their time is
		used for writing and executing tests, which was found in a study with 40 students~\cite{DBLP:conf/icse/BellerGZ15}. Since the study contained students,
		it may not generalize to companies, which is why almost 2,500 developers were monitored in companies for 2.5 years in a
		follow-up study, showing that developers spend a quarter of their work
		for testing---while believing they test half of their time
		\cite{DBLP:journals/tse/BellerGPPAZ19}. Our respondents appear to be
		slightly more realistic, estimating their testing at
		40\,\% or less.
		
		Developers would like to execute their tests more frequently
		but are handicapped by difficult testing frameworks as
		well as too little time given by management. This insight and that
		there is only a weak correlation between writing code and executing
		tests were found during a study with
		subsequent interviews~\cite{DBLP:conf/icsm/BlondeauEACCD17}. While this study focused on executing rather
		than writing tests, their conclusions about testing frameworks,
		conditions, and lack of time are confirmed by our survey.

		A recent survey on unit testing
		practices~\cite{DBLP:conf/issre/DakaF14} found that the developers
		are primarily driven by their conviction to test and management
		requirements, which matches our results. In addition, these survey results show that developers focus on writing, refactoring, and
		fixing code instead of writing tests because they do not enjoy
		testing, which can also be seen in the answers of our participants. In
		contrast to this work, we do not focus on unit tests, but on testing
		in general.
		
		According to a survey about thoughts on the career of a software
		tester~\cite{DBLP:journals/corr/WaychalC16},
		both students and professional testers think that testing is an
		important part of the software development cycle, but also tedious,
		frustrating and that they are missing developing software. Another study \cite{DBLP:journals/infsof/DeakSS16} also came to the same
		conclusion that missing recognition is one of
		the main reasons why testers are not satisfied with their job. In this
		work, we set our focus on the aspect of developer testing, but our
		findings are in agreement with these findings on software testers.
		
		Developers and testers in Brazilian companies have been reported to lack both training and knowledge in fundamental testing concepts \cite{DBLP:journals/jserd/MeloSSS22}.
		Moreover, testing itself apparently is generally not viewed as an important and prioritized 
		activity in the Brazilian companies involved in this study, which aligns
		with our global findings.
		
		\section{Conclusions} \label{conclusion}
		
		Insufficient testing is known to affect software quality. In order to
		better understand whether developers do not engage with
		testing because of technical, organizational, or motivational
		challenges, we conducted a survey with 284 participants. We find
		evidence that all three factors inhibit testing, and the details of how these factors affect testing can inform future research on how to improve testing practices.
		
		Of the many reasons that inhibit effective testing, we found
		evidence of a lack of intrinsic and extrinsic motivation in
		developers. Consequently, one potential opportunity to
		motivate developers to write more tests would be the
		application of external recognition systems, such as
		gamification, i.e., the inclusion of game elements to non-game
		related tools and
		contexts~\cite{DBLP:conf/mindtrek/DeterdingDKN11}. Such tools
		and approaches have been shown to provide benefits towards the
		motivation of developers in practice and educational scenarios
		to better engage with software
		testing~\cite{fulcini2023review}.
		
		To increase the generalization of the finding of this paper, it would
		be useful to replicate the survey with different audiences. To support
		replication, we provide a replication package containing all data and
		information:
		\url{https://doi.org/10.6084/m9.figshare.23212562}
		
		\section*{Acknowledgments}
		
		We would like to thank Marco Kuhrmann for his support while finding the right questions for our survey as well as all colleagues at the Chair of Software Engineering II at the University of Passau for their valuable input. This work is supported by the DFG under grant \mbox{FR 2955/2-1}, ``QuestWare: Gamifying the Quest for Software Tests''.

		\bibliographystyle{ieeetr}
		\bibliography{bib}
		
	\end{document}